\documentclass[3p,preprint, twocolumn]{elsarticle}
\usepackage{amsmath}
\usepackage{graphicx}
\usepackage{epstopdf}
\usepackage{epsfig}
\usepackage{tabu}
\usepackage{dcolumn}
\usepackage{bm}
\usepackage{color}
\usepackage{subfigure}
\usepackage{wrapfig}
\usepackage{rotate,color,url}
\usepackage{longtable}
\usepackage{time}
\usepackage{ulem}
\usepackage[pagebackref=false,colorlinks,linkcolor=blue,citecolor=blue,urlcolor=blue]{hyperref}
\usepackage{multirow,varwidth}
\usepackage{subfigure}

\newcommand{\Co}{Co$_2$Ni$_2$Nb$_2$O$_9$}
\newcommand{\CoFM}{Co$_2$Ni$_2$Nb$_2$O$_9$(FM)}
\newcommand{\CoAF}{Co$_2$Ni$_2$Nb$_2$O$_9$(AF)}
\newcommand{\NaCa}{NaCaNi$_2$F$_7$}
\newcommand{\MnFe}{MnFe$_2$O$_4$}
\newcommand{\Ca}{Ca$_{8.63}$Sb$_{10}$Sr$_{2.37}$}
\newcommand{\CrMn}{Cr$_{0.2}$Mn$_{0.6}$Ni$_{1.2}$Zr}
\DeclareUnicodeCharacter{0308}{HERE!HERE!} 
\begin{document}
\title{Machine Learning for compositional disorder: A Comparison Between Different Descriptors and Machine Learning Framworks}

\author[a]{Mostafa Yaghoobi}
\author[a]{Mojtaba Alaei\corref{author}}

\cortext[author] {Corresponding author.\\\textit{E-mail address:}  m.alaei@iut.ac.ir}
\address[a]{Department of Physics, Isfahan University of Technology, Isfahan 84156-83111, Iran}

\begin{abstract}
Compositional disorder is common in crystal compounds. In these compounds, 
some atoms are randomly  distributed at some crystallographic sites. For such compounds, 
randomness forms many non-identical independent structures.  
Thus, calculating the energy of all structures using ordinary quantum ab initio methods can be significantly time-consuming. 
Machine learning can be a reliable alternative to ab initio methods.  
We calculate the energy of these compounds with an accuracy close 
to that of density functional theory calculations in a considerably shorter time using machine learning.  
In this study, we use kernel ridge regression and neural network to predict energy. 
In the KRR, we employ sine matrix, Ewald sum matrix, SOAP, ACSF, and MBTR.
To implement the neural network, we use two important classes of application of the neural network in material science, 
including high-dimensional neural network and convolutional neural network based on crystal graph representation.
We show that kernel ridge regression using MBTR and neural network using ACSF can provide better accuracy than other methods.
\end{abstract}
\maketitle
\section{Introduction}
There is a common type of irregularity in the atomic structure of some crystalline compounds, 
called compositional disorder~\cite{Madelung1978}.
In these compounds, some crystallographic sites are occupied randomly by certain atoms 
, such that the chemical composition remains unchanged.
In a Crystallography Open Database~\cite{COD} (COD) survey, 
we find that approximately 31\% of entries have fractional site occupancies. 
Therefore, finding efficient computational solutions to deal with compositional disorders has high
priority in material science.
While the definition of compositional disorder is easily feasible through fractional site occupancies, 
simulation of such structures using {\it ab initio} methods such as density functional theory (DFT)  remains a challenge.

There are two prevalent {\it ab initio} approximations to deal with the compositional disorder.  
The combination of coherent potential approximation (CPA) ~\cite{Soven1967, Taylor1967} with a
band structure approach, the KKR - Kohn-Korringa-Rostoker - for muffin-tin potentials~\cite{Korringa1947, kohn1954}, provides an ab initio approach for randomly crystalline systems. 
The CPA-KKR~\cite{Stocks1980} is based on Green's function, which is not compatible with the architectures of many DFT codes. 
Another approximation is to use a certain cell of the disorder system and consider all possible combinations~\cite{Grau_Crespo2007}.
The bottleneck of this approach is the number of possible structures.
For example, due to the random distribution of Ca/Pr and N/O atoms at some Wyckoff sites in Pr$_{4-x}$Ca$_{x}$Si$_{12}$O$_{3+x}$N$_{18-x}$, 
with x=1.5, 34864 independent structures are found\cite{hong2017density}. 
Therefore applying DFT to consider all possible structures for many compositional disordered compounds (CDCs) are not practical.  

In the present work, we seek to calculate the energy of CDCs in a much shorter time 
than DFT without losing the accuracy of the calculations. 
Machine learning (ML)\cite{jordan2015machine} is one of 
the effective ways to replace DFT due to its ability to reduce computational costs significantly\cite{faber2017machine}.
Today, ML has been spread out as an interdisciplinary major in various fields.
The application of ML in chemistry, physics, and computational material science has been accompanied by crucial advances in recent years due to its speed and efficiency.

ML is a data-driven science\cite{solomatine2009data} and can be used to solve various problems if sufficient quality data exist.
In condensed matter physics, 
the first step to employ ML methods to predict material properties is to create appropriate features as ML input, 
the so-called descriptors. 
There are two main approaches to constructing material descriptors. 
The first approach uses only stoichiometry information.
For example, Ward et al.\cite{ward2016general} have proposed the general-purpose material descriptor, including stoichiometry, electronic structure, ionic compound attributes, 
and elemental properties statistic to create suitable inputs. Goodall et al.\cite{goodall2020predicting}, 
based on the vector representation of constituent elements of the atomic systems and the correlation between elements, 
introduced the stoichiometry graph as a material descriptor. Dipendra Jha et al.\cite{jha2018elemnet} 
have used only the elemental composition of materials as inputs for the deep neural network. 
However, none of these methods is suitable for predicting CDCs' properties due to the fixed stoichiometry in the structure. 
Therefore, we should use the second approach, which is based on crystallographic information.
There are many descriptors based on structural and geometrical information, including the Coulomb matrix\cite{rupp2012fast}, permutation invariant polynomials\cite{jiang2013permutation}, 
property-labeled material fragments\cite{oses2017universal}, Gaussian radial distribution functions\cite{seko2017representation}, bag-of-bands\cite{hansen2015machine}, 
angular Fourier series\cite{schmidt2019recent,seko2017representation}, and so forth. 
Here we use atom-centered symmetry function (ACSF)\cite{behler2011atom}, many-body tensor representation (MBTR)\cite{huo2017unified}, 
smooth overlap of atomic position (SOAP)\cite{willatt2019atom,bartok2013representing,imbalzano2018automatic}, 
Ewald sum matrix, and sine matrix (SM) \cite{faber2015crystal} to represent atomic structures.

After making the appropriate structural features for the ML, based on descriptors, 
we need to train a model to predict the total energy of systems. 
These models are mainly based on kernel ridge regression (KRR)\cite{zhang2013divide,vovk2013kernel,welling2013kernel}, 
Gaussian process regression\cite{quinonero2005unifying,bartok2015g}, 
and neural network (NN)\cite{fine2006feedforward,svozil1997introduction}. 
Tian Xie\cite{PhysRevLettXie} recently introduced another type of ML framework 
called crystal graph convolutional neural network (CGCNN) based on graph representation 
to learn the properties of materials directly from the graph created for crystals. 
This method eliminates the need to manually construct feature vectors for samples and the complexity behind it.

In the CDCs, the crystal sites and chemical formula remain unchanged for all possible combinations. 
Therefore, these materials are good platforms for benchmarking the ML models. 
In other words, we expect that a standard ML model should work exceptionally well for such materials. 
If a ML model cannot successfully describe the properties of these materials, 
we expect that it will not work correctly in other atomic environments.
In this study, we provide a small benchmark for descriptors and models for predicting total energy for 
the CDCs.

The paper is organized as follows: In section \ref{sec1}, we explain the computational details, 
which include structural information and the DFT calculation details, as well as a summary of the descriptors, 
and clarify the ML methods, including KRR and NN. 
In Section \ref{sec2}, we present the results, 
including the prediction and evaluation of the energy, 
and the error of the prediction for each method, then we compare the different methods.
\section{Computational Details}\label{sec1}
\subsection{Dataset details}
\begin{figure}
\centering
\includegraphics[scale=0.28]{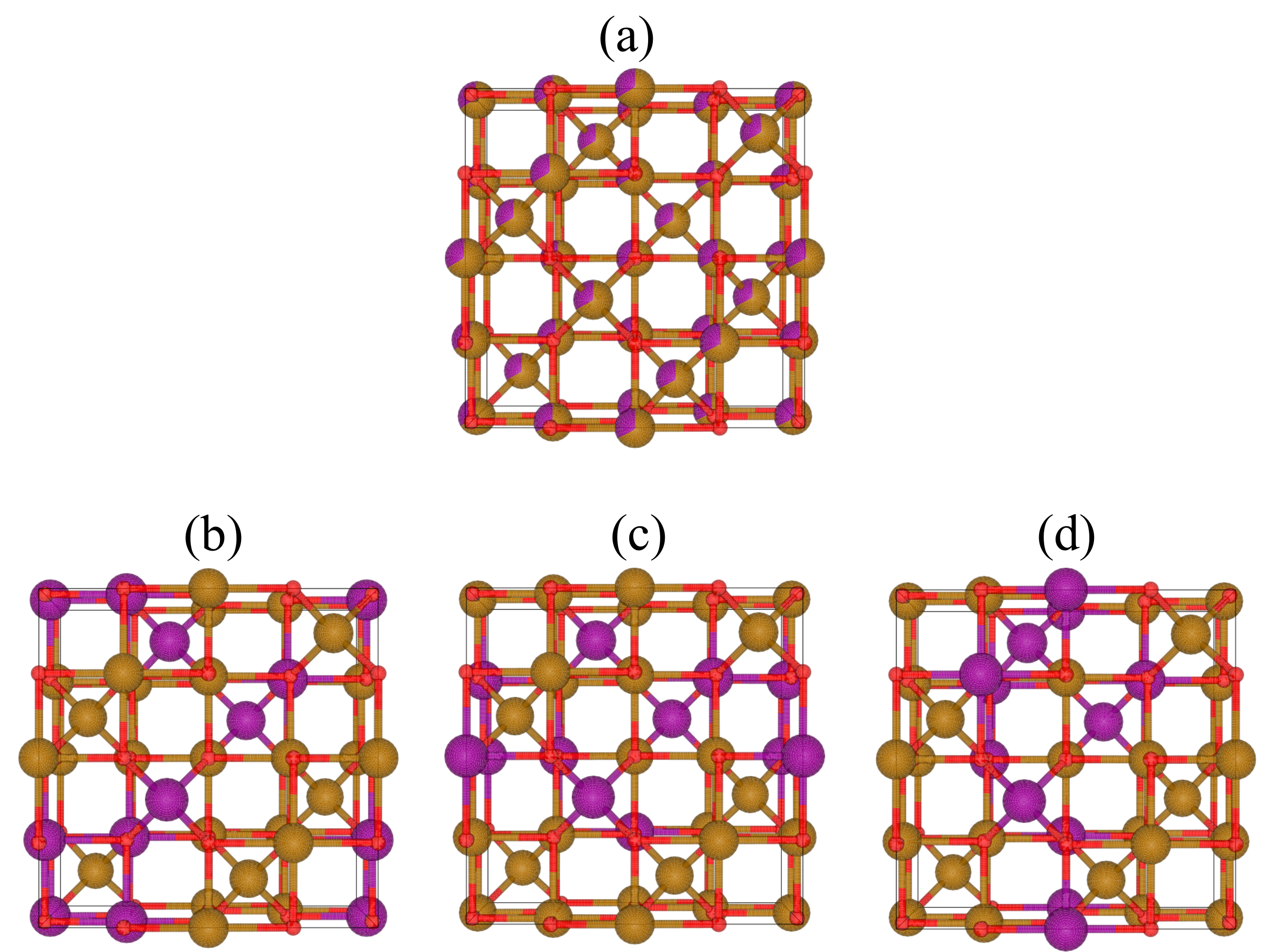}
\caption{Atomic structure of \MnFe. Iron, Manganese, and Oxygen atoms are shown in dark goldenrod, purple, and red, respectively. There are 1337 independent structures of this compound. 
         a) A schematic of the crystal structure of the conventional unit cell of \MnFe, with a Fe/Mn disorder in 8a and 16d Wyckoff positions with a probability of 0.667/0.333, 
         b) represents the most stable structure energitcally, that is, a structure with the lowest total energy, 
         c) represents the structure whose total energy indicates the median value among 1337 structures.
         d) represents the most unstable structure, i.e., the structure with the highest energy among all structures.}
\label{fig:Mnfe2o4}
\end{figure}

The dataset under study consists of five CDCs, including 
\NaCa\cite{krizan2015nacan}, \MnFe\cite{solano2014neutron}, 
\Ca\cite{gupta2006preferential}, \CrMn\cite{moriwaki1991electrode}, and \Co\cite{Papi2019magnetic}.
For \Co, we consider ferrimagnetic and antiferromagnetic configurations.
We name the ferrimagnetic  and antiferromagnetic phases of \Co~ with \CoFM~ and \CoAF, respectively. 
The structural information for each material is provided in the supplementary information.
As mentioned, due to the fractional occupancies of some sites in these materials, considerable independent structures may be found for each compound. 
Therefore, we use the Supercell program\cite{okhotnikov2016supercell} to obtain independent atomic configurations for each compound. 
The Supercell program first creates all atomic configurations for a compound inside a cell and then recognizes the atomic configurations, matched together by symmetric operations.
The results are symmetry-independent configurations. For instance, the total number of possible compounds related to \MnFe~ will be equal to $244608$, of which only 1337 structures are unique and independent.
Using the Supercell program, we find 97, 317, 644, 1337, and 280 independent structures for \NaCa, \Ca, \Co, \MnFe, and \CrMn, respectively. 
For \Co, we obtain the DFT data for both antiferromagnetic and ferrimagnetic phases.
We illustrate the several structures of \MnFe~ in figure \ref{fig:Mnfe2o4} 
as a representative of compositional disorder in this study to clarify this kind of structural irregularity.
We uploaded all DFT input and output files of  independent structures for these compounds in the novel materials discovery
(NOMAD) repository~\cite{CrMn,NaCa,CoNi_af,CaSb,CoNi_fm,MnFe}.

\subsection{DFT calculation}
After obtaining independent structures, we use DFT using the Quantum Espresso code (QE)\cite{giannozzi2009quantum} to calculate the total energy for each structure.
All compounds, except \Ca, are magnetic materials. 
So we use spin-polarized DFT calculations for the magnetic compounds and non-polarized DFT calculations for \Ca. 
The generalized gradient approximation (GGA) introduced by Perdew, Burke, and Ernzerhof (PBE) with improvement for solid structure (PBEsol)\cite{perdew2008restoring} 
is employed as an exchange-correlation potential. For all materials, we use the SCF calculation, along with a kinetic cutoff energy of 40 Ry.  
For all materials except for \CoAF~ and \CoFM, 
we set the k-point mesh to 4$\times$4$\times$4 using the Monkhorst–Pack scheme, but for \CoAF~ and \CoFM~ we use 3$\times$5$\times$1 k-mesh.
\subsection{Representation}
As mentioned, because of the special characteristic of the CDCs, 
we investigate the total energy of these materials based on the atomic representations containing structural information. 
Such representations are divided into global (i.e., describe the entire system by relationships between each atom) 
and local (i.e., describe the environment around every atom in the system). 
We use both global representations (sine matrix, Ewald sum matrix, and MBTR) and local representation (SOAP and ACSF).  
In the following, we provide a brief explanation of these descriptors. For details of the descriptors, see the supplementary information.
	
The Ewald sum matrix~\cite{faber2015crystal} is an extension of the Coulomb matrix for periodic systems. 
The same basic idea of the Ewald technique and the assumption of the neutral charge of the system are used to construct the Ewald sum matrix. 
It starts from the Coulomb potential, but considers all the infinite repetitions of the atoms along three orientations of the lattice vectors. 
Finally, the electrostatic potential between atoms is broken down into three terms: short-range and long-range potentials, and a constant term. 
The sum of these three terms for each atom in the system constructs the elements of the Ewald sum matrix.
	
The sine matrix~\cite{faber2015crystal} is another way to generalize the Coulomb matrix to periodic systems. 
The creative idea of this descriptor is to replace long-range electrostatic interactions with an arbitrarily constructed two-body potential. 
Although this potential has no physical meaning, it captures several physical properties of the system, 
such as the periodicity and the repetition of the potential contribution of the same atoms 
throughout the system and the infinite potential between two atoms that are immediately close to each other.

The basic idea of the MBTR~\cite{huo2017unified} is based on bag-of-bands~\cite{hansen2015machine}. 
The MBTR provides a measurement of the system's geometrical (distance and angle) and chemical (atomic numbers) properties
using k-body functions. Typically, k-body functions for k=1,2,3 encode atom types, 
distance or inverse distance between all pairs of atoms, and angle or cosine angle distributions for any triple of atoms, respectively. 
Although there are no restrictions on using higher-order k-body terms, 
it seems that up to the 3-body terms, this descriptor can be an acceptable representative of atomic systems. 
Each k-body is broadened into a Gaussian distribution to create a continuous numerical set for each element. 
To create the final MBTR vector, all different distributions of the k-body terms concatenate to each other \cite{huo2017unified}.

Another type of representation for atomic structures is the SOAP~\cite{willatt2019atom,bartok2013representing,imbalzano2018automatic}, 
which starts with constructing the Gaussian atomic density for each atomic environment. 
As mentioned previously, a descriptor must be invariant with respect to the rotation of the system. 
Therefore, the atomic densities are expanded using radial basis functions and spherical harmonic functions.
The SOAP's elements are obtained by multiplying the expansion coefficients for different atomic environments, known as the power spectrum. 
For more information, refer to Refs.\cite{caro2019optimizing,willatt2019atom,bartok2013representing,imbalzano2018automatic}.

In the ACSF~\cite{behler2007}, after generating atomic environments, the next step is to describe the positions of neighboring atoms inside the cut-off sphere. 
To do this, two types of functions, radial and angular symmetric functions introduced by Behler and Parrinello\cite{behler2007}, are used. 
The radial type is the sum of products of Gaussian densities and cut-off functions for all atoms inside the cut-off sphere.
The angular type is the sum of the cosine functions of the angles between all the triple atoms inside the cut-off sphere. 
The angular part can be multiplied by a Gaussian function to adjust according to the scale of the atomic distances.

A molecule or crystal can be represented by a graph. 
In CGCNN, each graph (for both molecules and crystals) contains nodes representing atoms 
and edges representing the connections between atoms in the structure.
Connections are not necessarily chemical bonds but can be any interaction between two atoms. 
Each node $i$ and  edge $(i,j)_k$ are embedded in a graph as feature vectors $\boldsymbol{v}_i$ and $\boldsymbol{u}(i,j)_k$, respectively. 
Feature vector $\boldsymbol{u}(i,j)_k$ represents $k$th connection between atom $i$ and $j$. 

Here, we use the python-based computational package "DScribe"\cite{himanen2020dscribe} to map atomic structures into the descriptors for use in the KRR method.
We use the "Atomic Simulation Environment" (ASE)\cite{larsen2017atomic} computational python package to supply atomic structures to DScribe.
\subsection{ML Methods}
\subsubsection{Kernel Ridge Regression (KRR)}
Kernel methods are an essential part of machine learning algorithms. 
KRR is based on ridge regression\cite{marquardt1975ridge}, 
in which the kernel method is used. The kernel method uses a trick, in which case the data is transferred to a higher-dimensional space 
by a function called the kernel function. Briefly, the energy predicted by machine learning can be obtained for compound $j$ in the test set as follows:
\begin{equation}
E^{\mathit{ML}}(\tilde{\boldsymbol{x}}_j)=\sum\limits_{i=1}^{n} \alpha_ik(\boldsymbol{x_i},\tilde{\boldsymbol{x}}_j),
\end{equation}
where $\tilde{\boldsymbol{x}}_j$ is the descriptor of the compound j, which we want to predict its energy, 
and $\boldsymbol{x_1},\boldsymbol{x_2},...,\boldsymbol{x_n}$ are the descriptors of the $n$ training compounds.
The regression coefficients $\boldsymbol{\alpha}$ are obtained by minimizing the regularized loss function.
Here $k(\boldsymbol{x_i},\tilde{\boldsymbol{x}}_j)$ represents the kernel function. In this study, we use Gaussian kernel defined as follows:
\begin{equation}
k(\boldsymbol{x_i},\tilde{\boldsymbol{x}}_j)=\exp\left(\frac{-||\boldsymbol{x_i}-\tilde{\boldsymbol{x_j}}||_2^2}{2\sigma^2}\right),
\end{equation}
where the hyperparameter $\sigma$ controls the Gaussian width\cite{rupp2015machine}.
The regression coefficients $\boldsymbol{\alpha}$ are obtained by minimizing the regularized loss function. 
The $\boldsymbol{\alpha}$ matrix is defined as $(K+\lambda I_{n\times n})^{-1} E$, 
where $K_{ij}=k(\boldsymbol{x_i},\boldsymbol{x_j})$ and $\lambda$ is the regularization hyperparameter. 
$I_{n\times n}$ denotes the identity matrix, and $E$ is a matrix that contains the DFT energies of the training compounds.
\subsubsection{High-dimentional Neural Network}
In this work, we use the high-dimensional neural network potential (HDNNP) introduced by Behler\cite{behler2011atom,behler2015constructing}. 
In HDNNPs, the ACSFs are first constructed for each structure. 
Then a separate NN is modeled for each atomic environment to predict the individual atomic contribution property 
(e.g., energy) associated with each atomic environment. 
Finally, to calculate the energy of the structure, 
we must sum all the contributions of each atomic environment obtained via each NN, 
as the following equation:
\begin{equation}
E=\sum_{i}E_i,
\end{equation}
here $E$ represents the total energy, and $E_i$ indicates the contribution energy of each atomic environment. 
We use the RuNNer package\cite{behler2017first} to predict the total energy through the HDNNP. 
This package is a reliable set for implementing HDNNPs. 
Using this package, the ACSFs are generated for each atomic system (periodic and non-periodic), 
and then the energy of each structure is obtained by implementing the aforementioned  algorithm\cite{behler2015constructing}.
\subsubsection{Crystal Graph Convolutional Neural Network (CGCNN)}
CGCNN is a deep learning framework for predicting the properties of crystal structures represented by crystal graphs. Once a graph is constructed from a structure, the convolutional layers repetitively update the atom feature vector $\boldsymbol{v}_i$ based on its neighbor atoms and related bonds:
\begin{equation}
\boldsymbol{v}_i^{t+1}=Conv(\boldsymbol{v}_i^{(t)},\boldsymbol{v}_j^{(t)},\boldsymbol{u}_{(i,j)_k}),
\end{equation}
here $Conv$ stands for convolution function. 
Then the pooling function is used to create an overall feature vector $\boldsymbol{v}$, which satisfies permutation invariance. 
In addition, there are two fully-connected layers, and eventually, an output layer is used to predict the target property.
\subsubsection{Hyperparameter Tuning}
We obtain the optimal values of hyperparameters, including  $A$ (accuracy parameter) and $W$, 
in the Ewald, and $r_{cut}$, $l_{max}$ and $n_{max}$, as well as the type of radial function (polynomial or Gaussian-type orbital) in the SOAP, 
and the $\sigma_k$, $s_k$, and $w_{k}^{min}$ values in the MBTR, 
as well as $\lambda$ and $\sigma$ in the KRR, through the grid search process. 
We also set the maximum number of neighbors for each node, maximum distance between two distinct atoms, 
number of convolution, and the depth of fully-connected layer, in CGCNN, through the grid search process.
We use the same method to find $\eta$ in radial the ACSF in the KRR, but in the NN, we use RuNNer code  instruction. 
Details related to each can be found in the Supplementary.
\section{Results}\label{sec2}
\begin{table}[th!]
\centering
\caption{\small Values for average RMSE, MAE and, $R^2$ parameter for the all models trained on 25 percent of data. 
Here Ewald, MBTR, SOAP, ACSF,and SM are the descriptors used in the KRR, and HDNNP refers to the high-dimensional neural network model using the RuNNer code, 
and CGCNN refers to crystal graph convolutional neural network.}
\resizebox{\columnwidth}{!}{%
\begin{tabular}{|c|c|c|c|c|}
\hline
compound&descriptor&RMSE(mHa)&MAE(mHa)&$R^2$\\\hline
\multirow{7}{*}{\NaCa}&SM&16.268&12.167&0.66083\\&Ewald&17.011&13.347&0.59225\\&SOAP&1.852&1.316&0.99547\\&ACSF&13.617&10.561&0.74229\\&MBTR&0.737&\bf{0.439}&0.99918\\&HDNNP&\bf{0.641}&0.529&\bf{0.99972}\\&CGCNN&2.343&1.389&0.99149\\
\hline
\multirow{7}{*}{\CrMn}&SM&2.345&1.865&0.57907\\&Ewald&2.500&1.979&0.55206\\&SOAP&2.086&1.676&0.67021\\&ACSF&2.513&1.989&0.54895\\&MBTR&1.169&0.903&0.89200\\&HDNNP&\bf{1.066}&\bf{0.862}&\bf{0.91016}\\&CGCNN&2.177&1.721&0.65141\\
\hline
\multirow{7}{*}{\CoAF}&SM&2.992&2.229&0.59007\\&Ewald&0.199&0.157&0.99811\\&SOAP&0.201&0.159&0.99807\\&ACSF&0.200&0.158&0.99808\\&MBTR&\bf{0.074}&\bf{0.057}&\bf{0.99973}\\&HDNNP&0.111&0.089&0.99944\\&CGCNN&0.178&0.133&0.99851\\
\hline
\multirow{7}{*}{\CoFM}&SM&1.216&0.963&0.86489\\&Ewald&0.420&0.312&0.98268\\&SOAP&0.184&0.144&0.99681\\&ACSF&0.195&0.143&0.99601\\&MBTR&\bf{0.127}&\bf{0.101}&\bf{0.99843}\\&HDNNP&0.133&0.104&0.99761\\&CGCNN&0.189&0.148&0.99672\\
\hline
\multirow{7}{*}{\MnFe}&SM&0.597&0.465&0.94467\\&Ewald&0.312&0.240&0.98461\\&SOAP&0.252&0.199&0.98997\\&ACSF&0.225&0.159&0.99208\\&MBTR&0.095&0.074&0.99856\\&HDNNP&\bf{0.089}&\bf{0.069}&\bf{0.99883}\\&CGCNN&0.339&0.269&0.98213\\
\hline
\multirow{7}{*}{\Ca}&SM&0.045&0.031&0.96214\\&Ewald&0.061&0.043&0.93058\\&SOAP&0.002&0.001(6)&0.99991\\&ACSF&0.041&0.031&0.96481\\&MBTR&\bf{0.001(5)}&\bf{0.001(2)}&\bf{0.99995}\\&HDNNP&0.009&0.007&0.99852\\&CGCNN&0.015&0.008&0.99568\\
\hline
\end{tabular}}
\label{tab2error}
\end{table}
\begin{figure*}[ht!]
\centering
\includegraphics[scale=0.38]{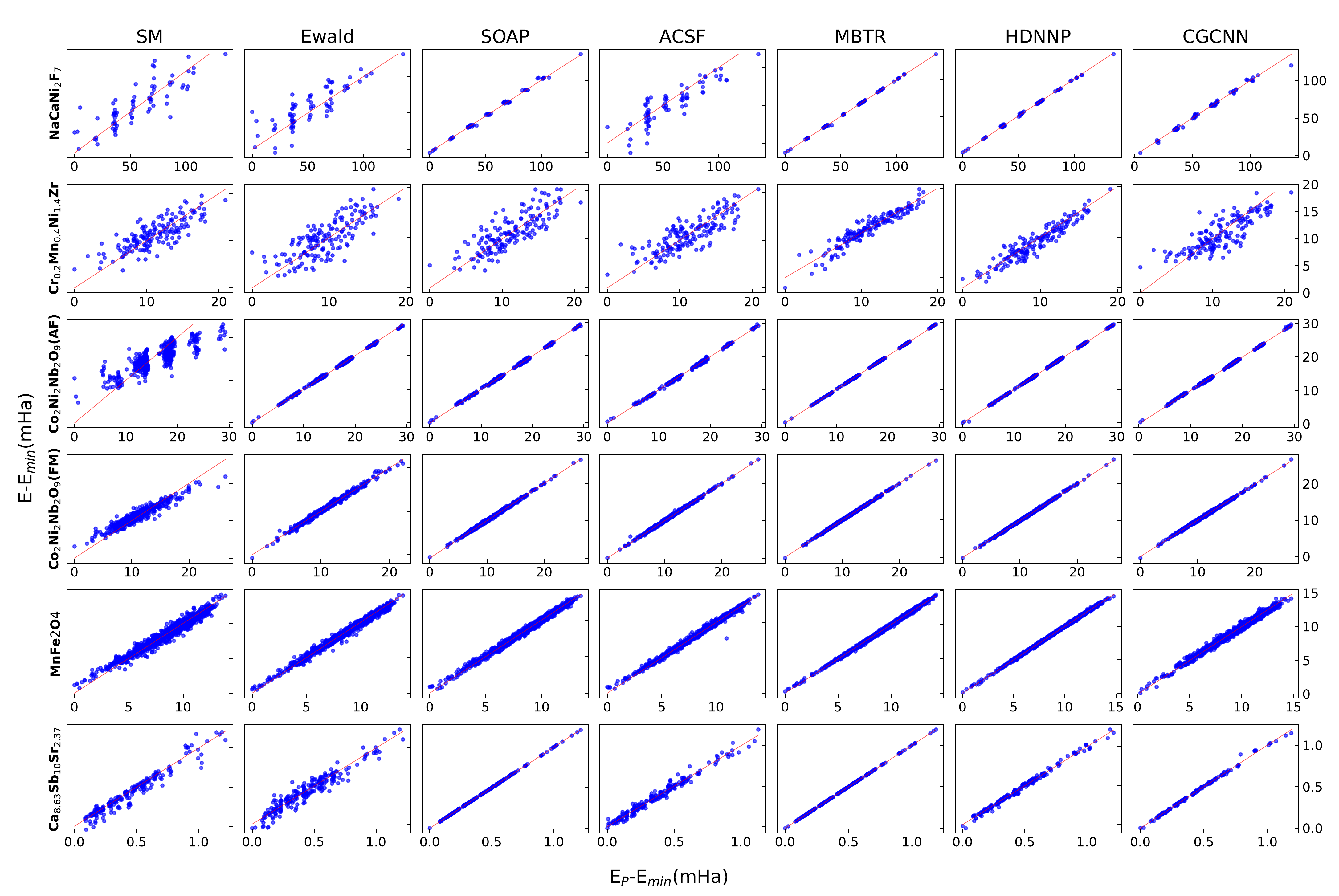}
\caption{\small The first to fifth columns show the models obtained through KRR using sine matrix, 
Ewald sum matrix, SOAP, ACSF, and MBTR. 
The sixth column shows the models obtained via HDNNP using RuNNer for all structures. 
All models are trained on 25\% of data, and 75\% of all data hold out of the training set.}
\label{KRR&RuNNer}
\end{figure*}
The scatter plots of the ML and DFT total energies obtained through all models, corresponding to each structure, 
are shown in Figure \ref{KRR&RuNNer}. 
To train all the models, we use 25\% of the data related to each compound. 
The y-axis shows the difference between the energy predicted by the ML model and the lowest DFT total energy. 
The x-axis shows the difference between the energy obtained through DFT and the DFT lowest total energy. 
According to this Figure, KRR+MBTR and HDNNP indicate the best performance.
	
To check the accuracy and assess the models, we use the root mean square error (RMSE)~\cite{geron2019hands}, 
the mean absolute error (MAE)~\cite{geron2019hands}, and the R-squared parameter ($R^2$)~\cite{raschka2015python}, 
defined in the following equations:
\begin{equation}\label{RMSEEQ}
\textrm{RMSE}=\sqrt{\frac{1}{n}\sum_{i=1}^{n}(\hat{\varepsilon}^{(i)}-\varepsilon^{(i)})^2}
\end{equation}
\begin{equation}\label{MAEeq}
\textrm{MAE}=\frac{1}{n}\sum_{i=1}^{n}|\hat{\varepsilon}^{(i)}-\varepsilon^{(i)}|
\end{equation}
\begin{equation}\label{rsquaredeq}
R^2=1-\frac{\sum_{i}(\hat{\varepsilon}^{(i)}-\varepsilon^{(i)})^2}{\sum_{i}(\varepsilon^{(i)}-\bar{\varepsilon})^2},
\end{equation}
here $\hat{\varepsilon}^{(i)}$ is the total energy value predicted by the machine learning model for $i$th structure, 
$\varepsilon^{(i)}$ is the DFT total energy of the $i$th structure, and $\bar{\varepsilon}$ represents the average of DFT total energies ($1/n\sum_{i=1}^n \varepsilon^{(i)}$).
The closer the $R^2$ is to 1, the better the model has been trained and is more affected by features changes. 
	 
Table \ref{tab2error} shows the values of RMSE, MAE, and $R^2$ for the models that have been obtained through training on 25 percent of the structures of each compound.
According to this table, except for \Ca; the best results are obtained for  KRR+MBTR and HDNNP. For \Ca, 
KRR+SOAP and KRR+MBTR have better performance, although HDNNP values are very close to them.
KRR+SOAP almost presents the second-best results. 
In most cases, the KRR+ACSF also has similar KRR+SOAP results, except for \Ca~ and \NaCa.
The worst result of KRR+ACSF is associated with \NaCa~ with a very large RMSE ($\sim 13$ mHa). 
In most cases, the KRR+SM and KRR+Ewald exhibit the largest RMSE and MAE and lowest $R^2$. 
CGCNN gives the best results after KRR+SOAP, except for \CoAF~ and \MnFe. 
For \CoAF, CGCNN is the third accurate model. For \MnFe, CGCNN is the sixth most accurate model after Ewald.
In all cases, the RMSE and $R^2$ are compatible; therefore, in our study, $R^2$ does not add additional information about the quality of our training.

\begin{figure*}[t!]
	\centering
	\includegraphics[scale=0.29]{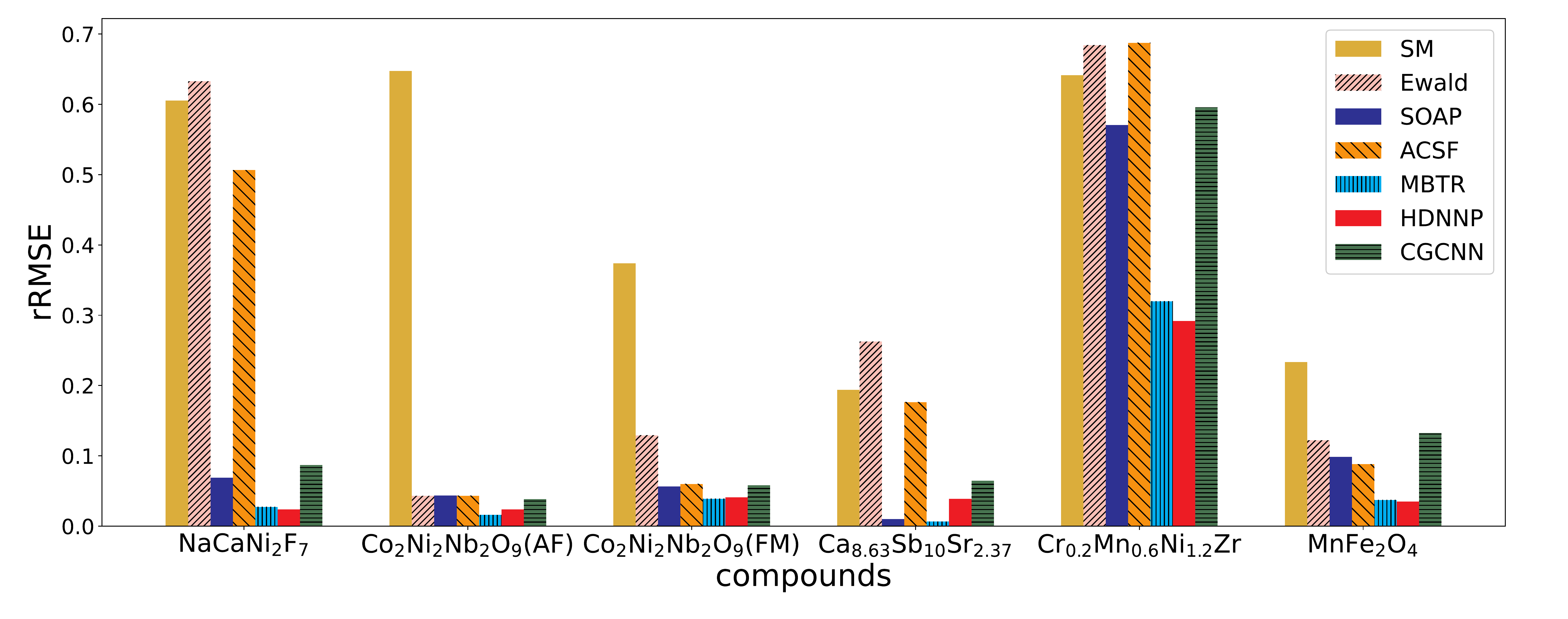}
	\caption{The bar plot for the values of the rRMSE for the all models. All models are trained on 25\% of the structures.}
	\label{figRMSE}
\end{figure*}

To show the error of the prediction values for all compounds in one graph, since RMSE is scale-dependent, 
we use the scale-independent relative RMSE (rRMSE)\cite{botchkarev2019new}, which is defined as follows:
\begin{equation}
\textrm{rRMSE}=\frac{\textrm{RMSE}}{\sqrt{\frac{1}{n}\sum\limits_{i=1}^{n}(\varepsilon^{(i)}-\bar{\varepsilon})^2}}=\frac{\textrm{RMSE}}{\sigma_{\varepsilon}},
\end{equation}
where $\bar{\varepsilon}=\frac{1}{n}\sum_{n}^{i=1}\varepsilon^{(i)}$ is the mean of the energies, and $\varepsilon^{(i)}$ is the energy of the $i$th structure. 
$\sigma_{\varepsilon}$ indicates the standard deviation of the energies. 
In Figure \ref{figRMSE}, we show the rRMSE of all compounds and methods. 
According to Figure \ref{figRMSE}, the maximum value of rRMSE for most cases is obtained via the KRR+SM and KRR+Ewald models.
The diagram indicates the lowest rRMSE values for KRR+MBTR and HDNNP.    
For all trained models on 25 percent of the structures through the KRR with the MBTR and SOAP and the HDNNP using the RuNNer, 
it is possible to find the stable structure (i.e., the structure with the lowest energy among the existing structures).
We exclude the stable structures from the training sets and hold the stable structures in the test sets deliberately
for benchmarking the models to check the capability of the ML methods
to find stable structures. 
The KRR+ACSF models could find stable structures for \CoAF, \CoFM, \Ca, and \MnFe, 
and the KRR+Ewald models could find stable structures only for \CoAF~ and \CoFM. 
CGCNN is able to find stable structure only for \CoAF, \CoFM, \NaCa, and \Ca.
We figure out it is impossible to find stable structures for these compounds through the KRR+SM, 
except in exceptional cases (for example, when we increase the number of data in the training set).
\begin{figure*}[t!]
\centering
\includegraphics[scale=0.35]{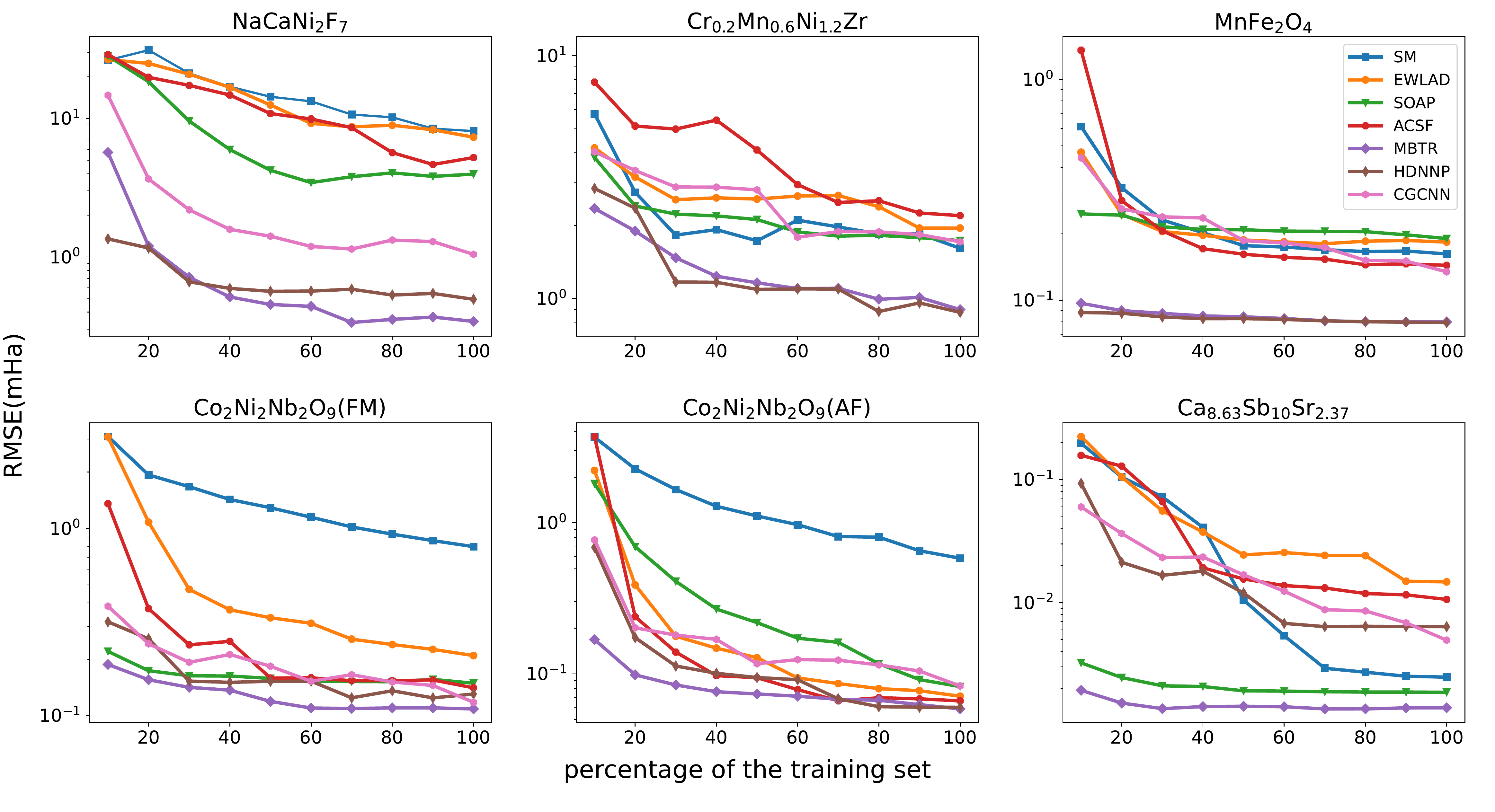}
\caption{The learning curve for RMSE, as a function of training set size, for all models.}
\label{figLC}
\end{figure*}
	
In Figure \ref{figLC}, we plot the learning curve\cite{anzanello2011learning} 
(a diagram showing the model error, considered here RMSE, as a function of the size of the training set) 
for all structures and all methods.  
At each step, we increase the number of structures in the training set by 10 percent. 
The data for training are taken from 80 percent of the total data. 
Then, the RMSE is reported for the remaining 20 percent of the total data.
In some cases of the learning curves, it is worth mentioning that using the optimized value of the hyperparameter $\lambda$ 
makes the $\mathbf{\alpha}$ matrix negative-definite. Therefore, for the learning curve of some compounds, 
such as \NaCa, we change the hyperparameter $\lambda$ to avoid the negative-definite problem. 
Changing $\lambda$ causes a different numerical result between Table \ref{tab2error} and the learning curve, but the trends in both are the same.

In this plot, for all methods and structures, the downward trend for the RMSE is visible as the training set size increases. 
It seems that by increasing the number of data in the training set to more than 60\%, the error values will not change substantially. 
The lowest error is obtained for all compounds through the KRR+MBTR and HDNNP, except for \Ca, 
where the KRR+SOAP performs better than the HDNNP.
According to this Figure, for \NaCa, only the HDNNP and KRR+MBTR models are well trained. After these methods, CGCNN also shows a reliable curve.
In other cases, even with 100\% of the data allocated to the training set (77 out of 97 structures), 
RMSE hardly reaches 5mHa, which indicates the weakness of these models.
For \CrMn, KRR+ACSF shows the worst graph among all others.
For \CoAF, the diagram related to the KRR+SM shows the worst forecasting model. The other diagrams show a good fit such that by increasing 
the training set size to more than 30\% (154 structures), 
they all converge to a value close to 0.1mHa.
For \CoFM, we achieve the best learning curves using the KRR+MBTR and SOAP, and HDNNP.
The worst diagram is related to the KRR+SM, which shows that this method is unreliable 
even if the training set increases to more than 60\%.
The KRR+Ewald curve is the worst model among other models after the KRR+SM, 
although it goes to 0.2 mHa after increasing the training set to more than 40\%.
As can be seen from the learning curves related to \MnFe, KRR+MBTR and HDNNP have the best performance over other methods. 
Other curves converge by almost the same amount after increasing the training set size to more than 30\%.
The diagrams for \Ca~ clearly show the effect of increasing the training set size on the model improvement.
The curves for KRR+MBTR and KRR+SOAP start from very small RMSE, 
which means that by increasing the training set size to more than 20\% of the training data, 
a suitable model can be achieved. The curves for HDNNP and CGCNN show almost the same behavior, 
and both models after the previous two models have a high ability to predict. 
The Ewald, ACSF, and SM curves show almost the same behavior, although the SM curve, 
after increasing the training set size to more than 50\%(127 structures), converges to a smaller value than the other two models.

\section{conclusion}
We performed an unbiased evaluation of the KRR using the MBTR, ACSF, SOAP, SM, and Ewald descriptors, and 
HDNNP  with the ASCF descriptor using RuNNer, and CGCNN.
We chose the CDCs as a benchmark for the atomic descriptors and ML methods. 
We expect that we can quickly rule out the descriptors that do not work correctly with such compounds.
We found that the KRR+MBTR and HDNNP are excellent even at a meager amount of training data.
Except for one case (\CrMn), CGCNN efficiently forecasts energy with a performance comparable to KRR+SOAP.
The KRR+SOAP, along with KRR+MBTR and HDNNP, can also be trustworthy for the CDCs, 
mainly due to its ability to correctly predict the structure with the lowest total energy. 
A recent study~\cite{rupp2021} also indicates the potential of the SOAP for alloy compounds. 
The KRR+SM and KRR+Ewald result in large RMSEs and require more data for training. 
The achievement of the KRR+ACSF is case-dependent, and like Ewald and SM, it requires more training data to reach a reasonable accuracy. 
In summary, for CDCs, 
we recommend the KRR+MBTR, HDNNP, and KRR+SOAP to predict the total energy and find the structure with the lowest total energy.

\section{Acknowledgments}
This work was supported by the Vice Chancellor
for Research Affairs of Isfahan University of Technology (IUT).

\section{Data Availability}
Some part of the data that supports the findings of this study are available within the article and its supplementary material. 
In addition, the DFT data that used in this study are openly available in the NOMAD repository. 
To access the DFT data, visit reference numbers~\cite{CrMn,NaCa,CoNi_af,CaSb,CoNi_fm,MnFe}.

	\section{supplementary of Machine Learning for compositional disorder: A Comparison Between Different
		Descriptors and Machine Learning Frameworks}
	
	
	\subsection{Structures Information}
	
	Structural information about each compound is shown in the table \ref{supercellprogram}. This table shows the type of atoms, the Wyckoff positions, and the occupancy number of the atoms at each site. 
	The number of symmetry-independent structures derived from each compound and the atomic positions are shown too.
	\begin{table*}[th!]
		\resizebox{\width}{!}{%
		\tiny 
		\begin{tabular}{|cccccccc|}
			\hline
			compound&atoms&Wyckoff&Occ&x&y&z&number of independent compound\\\hline
			\multirow{5}{*}{NaCaNi$_2$F$_7$}&Ca&16d&0.5&0.5&0.5&0.5&\multirow{5}{*}{97}\\&Na&16d&0.5&0.5&0.5&0.5&\\&F&8b&1&0.375&0.375&0.375&\\&F&48f&1&0.125&0.125&0.125&\\&Ni&16c&1&0.0&0.0&0.0&\\\hline
			\multirow{8}{*}{Co$_2$Ni$_2$Nb$_2$O$_9$}&8d&Ni/Co&0.5/0.5&0.33550&-0.00240&0.18952&\multirow{8}{*}{644}\\&8d&Ni/Co&0.5/0.5&0.16460&0.50750&0.00008&\\&Nb&8d&1&0.02250&-0.00240&0.35571&\\&O&4c&1&0.00000&0.28790&0.25000&\\&O&8d&1&0.16460&0.16550&0.42374&\\&O&8d&1&0.16570&0.17180&0.09530&\\&O&8d&1&0.35260&0.35160&0.25200&\\&O&8d&1&0.49570&0.16790&0.08380&\\\hline	
			\multirow{9}{*}{Ca$_{8.63}$Sb$_{10}$Sr$_{2.37}$}&Sr/Ca&16n&0.35/0.65&0.2522(2)&0.0000&0.1875(1)&\multirow{9}{*}{317}\\&Ca&16n&1&0.3398(2)&0.0000&0.3963(1)&\\&Ca&8h&1&0.1681(5)&0.1687(5)&0.0000&\\&Sb1&4d&1&0.5000&0.0000&0.2500&\\&Sb2&4e&1&0.0000&0.0000&0.1283(1)&\\&Sb3&16m&1&0.2076(1)&0.2076(1)&0.3236(1)&\\&Sb4&8i&1&0.3447(1)&0.0000&0.0000&\\&Sb5&8h&1&0.1257(1)&0.1257(1)&0.50000&\\&Sr&4e&1&0.0000&0.0000&0.3374(2)&\\\hline
			\multirow{3}{*}{MnFe$_2$O$_4$}&Fe/Mn&8a&0.66667/0.33333&0.1250&0.1250&0.1250&\multirow{3}{*}{1337}\\&Fe/Mn&16d&0.66667/0.33333&0.5000&0.5000&0.5000&\\&O&32e&1&0.2505&0.2505&0.2505&\\\hline		
			\multirow{2}{*}{Cr$_{0.2}$Mn$_{0.6}$Ni$_{1.2}$Zr}&Cr/Ni/Mn&16d&0.1/0.6/0.3&0.6250&0.6250&0.6250&\multirow{2}{*}{280}\\&Zr&8a&1&0.0000&0.0000&0.0000&\\\hline
		\end{tabular}}
		\label{supercellprogram}
	\end{table*}
	
	\subsection{representation}
	In First Principle calculations, it can be said that only by knowing the type of atoms of each structure and the dependent positions of each atom, the properties related to each structure can be achieved. However, these parameters are not suitable for machine learning models. Imagine that with just one rotation or translation of the system in space, atomic positions may change, which may distort the prediction of the machine learning model. For this reason, from the beginning of using machine learning to search for material properties, finding the suitable inputs has been controversial. Machine learning inputs to find material properties are called descriptor. A representation of atomic structures, that are invariant concerning rotation, transmission of system in space and permutation of identical atoms in system. A prestigious descriptor should also have a one-to-one relationship with atomic structures, because a numerical set may describe several structures.  Additionally, the calculations related to the generation of the descriptor should not take considerable time\cite{huo2017unified,faber2015crystal,jager2018machine}.
	\subsection{Ewald Sum Matrix}
	The Ewald sum matrix to represent periodic systems uses the Coulomb matrix expansion for the periodic system. In this descriptor, the contribution of the infinite repetition of atoms in three directions of lattice vectors in interactions between two atoms is considered,
	\begin{equation}
		\phi_{ij}=\sum_{\mathbf{n}} \frac{Z_iZ_j}{|\mathbf{R}_i-\mathbf{R}_j|+\mathbf{n}},
	\end{equation} 
	where the sum over $n$ is taken over the all lattice vectors. In the Ewald sum matrix, the Ewald summation technique and a neutral background charge\cite{lee2009ewald,hub2014quantifying} are used to converge this sum. The basic idea is to split this sum into two term and one constant that converge quickly: 
	\begin{equation}
		\phi_{ij}=\phi_{ij}^{d}+\phi_{ij}^{r}+\phi_{ij}^0,
	\end{equation}
	here, $\phi_{ij}^{d}$ and $\phi_{ij}^{r}$ represent short range interaction calculated in direct space and the long-range interaction calculated in reciprocal space, respectively. $\phi_{ij}^0$ is a constant corrective term.
	The Ewald sum matrix is then defined as follows:
	\begin{equation}
		\text{D}_{ij}^{\text{Ewald}}=
		\begin{cases}
			\phi_{ij}^{d}+\phi_{ij}^{r}+\phi_{ij}^0&\forall \hskip 1mm i=j\\
			2\left(\phi_{ij}^{d}+\phi_{ij}^{r}+\phi_{ij}^0\right)&\forall \hskip 1mm
			i\neq j
		\end{cases},
	\end{equation}
	where the terms are given by:
	\begin{equation}\label{real}
		\phi_{ij}^{d}=\frac{1}{2}Z_iZ_j\sum_{n^\prime}\frac{\text{erfc}(\alpha|R_i-R_j+n|)}{|R_i-R_i+n|},
	\end{equation}
	\begin{equation} \label{recip}
		\phi_{ij}^{r}=\frac{2\pi}{V}Z_iZ_j\sum_{G}\frac{e^{-|G|^2/(2\alpha)^2}}{|G|^2}\cos(G.(R_i-R_j)),
	\end{equation}
	\begin{equation}\label{consEW}
		\phi_{ij}^0=
		\begin{cases}
			-\frac{\alpha}{\sqrt{\pi}}Z_i^2-\frac{\pi}{2V\alpha^2}Z_i^2&\forall \hskip 1mm i=j\\
			-\frac{\pi}{2V\alpha^2}Z_iZ_j& \forall \hskip 1mm i\neq j    
		\end{cases}.
	\end{equation}
	
	The prime notation here means that when $n=0$, then $i=j$ are not considered. $G$ is the reciprocal space lattice vector, $\alpha$ is the screening parameter and only affects the rate of convergence in the Ewald method. There are several suggestions for how to set it, But a well-tested recommendation is,
	
	\begin{equation}\label{alphaEW}
		\alpha=\sqrt{\pi}\left(\frac{WN}{V^2}\right)^{1/6}
	\end{equation} 
	
	whare $N$ is the number of atoms in the unit cell, and $V$ is the volume of the cell. $W$ is a weight parameter that controls the relative computational cost. 
	To avoid long calculations and achieve convergence, the summations in Eqs.\ref{real} and
	\ref{recip} stop at the $n_{cut}$ and $G_{cut}$ respectively.
	
	The appropriate values for the hyperparameters required to construct this descriptor are shown in Table \ref{tab:ewH}. According to Eqs.\ref{real}, \ref{recip} and \ref{consEW}, we need to determine the appropriate values for $\alpha$, $n_{cut}$ and $G_{cut}$. The values for $\alpha$ for each compounds is obtained through Eq.\ref{alphaEW}. Suitable values for $n_{cut}$ and $G_{cut}$ to construct the Ewald matrix in the DScribe package can be adjusted both automatically and manually. To set the values of $n_{cut}$ and $G_{cut}$ automatically, we need to estimate the value of A, which is known as accuracy parameter. By specifying this parameter, the appropriate values for $n_{cut}$ and $G_{cut}$ can be determined according to the following equations;
	$$n_{cut}=\frac{\sqrt{-\ln A}}{\alpha}$$
	$$G_{cut}=2\alpha \sqrt{-\ln A}  $$
	$\alpha$ and only affects the rate of convergence in the Ewald method.
	Table \ref{tab:ewH} shows the hyperparameter values used to determine the elements of the Ewald sum matrix for each compounds.

	\begin{table}[h!]
		\centering
		\caption{\small Hyperparameters required to generate Ewald sum matrix for the studied compounds. Here Co$_2$Ni$_2$Nb$_2$O$_9$(AF) according to Co$_2$Ni$_2$Nb$_2$O$_9$ in Antiferromagnetic phase, and Co$_2$Ni$_2$Nb$_2$O$_9$(FM) represents its Ferrimagnetic phase. These parameters are obtained through gridsearch process that for Co$_2$Ni$_2$Nb$_2$O$_9$(FM), Co$_2$Ni$_2$Nb$_2$O$_9$(AF), MnFe$_2$O$_4$, and Ca$_{8.63}$Sb$_{10}$Sr$_{2.37}$ trained on 100 structures, and for Cr$_{0.2}$Mn$_{0.4}$Ni$_{1.2}$Zr trained on 90, and for NaCaNi$_2$F$_7$ trained on 40 structures.}
		\resizebox{\columnwidth}{!}{%
		\begin{tabular}{|ccccc|}
			\hline\hline
			compounds&W&A&$\lambda$&$\sigma$\\ \hline
			NaCaNi$_2$F$_7$&0.9&$4.27\times10^{-7}$&$5.88\times10^{-15}$&-1196.7596\\
			Co$_2$Ni$_2$Nb$_2$O$_9$(AF)&0.088&$\times10^{-11}$&$6.30\times10^{-15}$&-1809.0349\\
			Co$_2$Ni$_2$Nb$_2$O$_9$(FM)&0.088&$5\times10^{-11}$&$5.88\times10^{-14}$&-1809.4098\\
			Ca$_{8.63}$Sb$_{10}$Sr$_{2.37}$&0.02&$1.3182\times10^{-6}$&$3.16\times10^{-15}$&-2658.8758\\
			MnFe$_2$O$_4$&0.09&$2.51\times10^{-9}$&$2.99\times10^{-14}$&-840.2713\\
			Cr$_{0.2}$Mn$_{0.6}$Ni$_{1.2}$Zr&0.09&$7.9\times10^{-8}$&$1.66\times10^{-10}$&-439.4255\\\hline
		\end{tabular}}
		\label{tab:ewH}
	\end{table}
	\subsection{Sine Matrix}
	The Ewald sum matrix is a helpful descriptor for periodic atomic structures, but sometimes when the atomic system is remarkably large, it can have quite complicated and long calculations. The sine matrix would come in handy in these situations. The matrix elements are defined by
	\begin{equation}\label{eqsine}
		\text{D}_{ij}^{\text{sine}}=
		\begin{cases}
			0.5Z_i^{2.4}&\forall \hskip 1mm i=j\\
			\phi_{ij}& \forall \hskip 1mm i \neq j
		\end{cases},
	\end{equation}
	where 
	\begin{equation}
		\phi_{ij}=Z_iZ_j\|B.\sum_{k=\{x,y,z\}}\hat{e}_k \sin^2(\pi B^{-1}.(R_i-R_j))\|^{-1},
	\end{equation}
	where $\hat{e}_k$ is the coordinate unit vectors and $B$ is a matrix that consists of lattice vectors of the system. This potential may not have physical meaning, but it can extract useful information from a periodic system. And just like the Coulomb matrix as described in \cite{ramakrishnan2015electronic}, the diagonal elements are obtained by fitting with the potential energy of neutral atoms. Table \ref{tab:sinematrix} shows the hyperparameters $\lambda$ and $\sigma$ related to the using this descriptor along with KRR.
	\begin{table}[h!]
		\centering
		\caption{\small The hyperparameters required to generate sine matrix. These parameters are obtained through grid search process.}
		\begin{tabular}{|ccc|}
			\hline\hline
			compounds&$\lambda$&$\sigma$\\ \hline
			NaCaNi$_2$F$_7$&$5.88\times10^{-15}$&-1194.5165\\
			Co$_2$Ni$_2$Nb$_2$O$_9$(AF)&$2.39\times10^{-14}$&-1808.9875\\
			Co$_2$Ni$_2$Nb$_2$O$_9$(FM)&$5.95\times10^{-14}$&-1809.4098\\
			Ca$_{8.63}$Sb$_{10}$Sr$_{2.37}$&$3.16\times10^{-15}$&-2656.7858\\
			MnFe$_2$O$_4$&$1.65\times10^{-14}$&$-840.27134$\\
			Cr$_{0.2}$Mn$_{0.6}$Ni$_{1.2}$Zr&$5.34\times10^{-14}$&-683.8454\\\hline
		\end{tabular}
		\label{tab:sinematrix}
	\end{table}
	
	This descriptor has no hyperparameter to determine in advance. Table \ref{tab:sinematrix} shows the hyperparameters of the KRR method when using this descriptor, which is obtained through the grid search.
	\subsection{Smooth Overlap Of Atomic Position(SOAP)}
	In SOAP, the atomic system is first divided into atomic environments, i.e, each atom is first considered as the central atom, then a sphere with a cutoff radius around the central atom is formed, then the Gaussian density $\rho^Z$ of the element Z, corresponding to the center of each atomic environment is calculated: 
	\begin{equation}\label{densitysoap}
		\rho^Z(r)=\sum_{i}^{|Z|}e^{-\frac{1}{2\sigma^2}|r-R_i|^2}
	\end{equation}
	Here the index $i$ runs over atoms with the atomic number $Z$ to obtain the density corresponding to that element, and $\sigma$ controls the width of the gaussian. Atomic density can be expanded by spherical harmonics and a set of orthonormal radial base functions $g_n$ as:
	\begin{equation}
		\rho^Z(r)=\sum\limits_{nlm}c_{nlm}^Zg_n(r)Y_{lm}(\theta ,\phi)
	\end{equation}
	The coefficients of this expansion can be calculated as follows:
	\begin{equation}
		c_{nlm}^Z=\iiint_{R^3}dVg_n(r)Y_{lm}(\theta ,\phi )\rho^Z(r)
	\end{equation}
	The atimic densities of the different atomic environments are then multiplied by each other, which is equivalent to the multiplication of the $c_{nlm}$ of each atomic environment. The final output of the SOAP descriptor can be in the form of a power spectrum vector for each system, the elements of which will be as follows:
	\begin{equation}
		p_{nn^\prime l}^{Z_1,Z_2}=\pi\sqrt{\frac{8}{2l+1}}\sum_{m}(c_{nlm}^{Z_1})^*c_{n^\prime lm}^{Z_2}
	\end{equation}
	Spherical harmonics constitute a complete set of orthogonal functions that are suitable for the angular description of the atomic environments. There is no single choice for the radial degree of freedom, one suitable choice can consist of orthogonalized cubic and higher order polynomials\cite{bartok2013representing}, defined as
	\begin{equation}
		g_{n}(r)=\sum_{n^\prime=1}^{n_{max}}\beta_{nn^\prime}\phi_{n^\prime}(r)
	\end{equation}
	\begin{equation}
		\phi_n(r)=(r-r_{cut})^{n+2}
	\end{equation}
	and set of spherical primitive gaussian type orbitals $g_{nl}(r)$, defined as
	\begin{equation}
		g_{nl}(r)=\sum_{n^\prime=1}^{n_{max}}\beta_{nn^\prime l}\phi_{n^\prime l}(r)
	\end{equation}
	\begin{equation}
		\phi_{nl}(r)=r^le^{-\alpha_{nl}r^2}
	\end{equation}
	The Dscribe way to orthnormalizing weights $\beta_{nn^\prime l}$ is to use Löwdin orthogonalization\cite{lowdin1950}. Essential hyperparameters for generating the SOAP for each compound are shown in table \ref{tab:soapH}. 
	\begin{table}[h!]
		\centering
		\caption{\small Hyperparameters required to generate SOAP. These parameters are obtained through grid search process that for Co$_2$Ni$_2$Nb$_2$O$_9$(FM), Co$_2$Ni$_2$Nb$_2$O$_9$(AF), MnFe$_2$O$_4$, and Ca$_{8.63}$Sb$_{10}$Sr$_{2.37}$ trained on 100 structures, and for Cr$_{0.2}$Mn$_{0.4}$Ni$_{1.2}$Zr trained on 90, and for NaCaNi$_2$F$_7$ trained on 40 structures.}
		\resizebox{\columnwidth}{!}{%
			\begin{tabular}{|cccccccc|} \hline\hline
				&$r_{cut}(\text{\AA})$&$l_{max}$&$n_{max}$&$\sigma$&rbf&$\lambda$&$\sigma$\\ \hline
				NaCaNi$_2$F$_7$&10&9&15&$10^{-4}$&gto&$4.46\times10^{-15}$&-1195.2654\\
				Co$_2$Ni$_2$Nb$_2$O$_9$(AF)&17&7&17&$10^{-4}$&gto&$6.30\times10^{-14}$&-1808.7441\\
				Co$_2$Ni$_2$Nb$_2$O$_9$(FM)&15&8&12&$10^{-3}$&gto&$4.36\times10^{-14}$&-1809.5482\\
				Ca$_{8.63}$Sb$_{10}$Sr$_{2.37}$&10&9&15&$10^{-4}$&gto&$3.16\times10^{-15}$&-2658.8475\\
				MnFe$_2$O$_4$&11&14&9&0.00081&polynomial&$3.99\times10^{-14}$&-839.9874\\
				Cr$_{0.2}$Mn$_{0.6}$Ni$_{1.2}$Zr&8.5&4&9&0.00021&gto&$6.33\times10^{-14}$&-683.8274\\\hline
		\end{tabular}}
		\label{tab:soapH}
	\end{table}
	\subsection{Many Body Tensor Representation (MBTR)}
	Here we have used the special representation introduced by H.Huo and M.Rupp\cite{huo2017unified} as a representation of molecules and crystals for machine learning. Briefly, MBTR is a descriptor that describes an atomic structure by 1-body (corresponding to the atoms in the structure), 2-body (corresponding to the distances between atoms), 3-body (corresponding to the angles between three atoms in the structure), 4-body functions (corresponding to the dihedral angles), etc. However, it seems that this descriptor can provide high-grade accuracy with 1-body, 2-body, and 3-body functions. A set of geometric functions $g_k$ is applied to create k-body functions. These functions are not unique and there can be different choices for each function. Dscribe package\cite{himanen2020dscribe} uses atomic numbers for $g_1$, distance or inverse distance between atoms for $g_2$, and angle or cosine angle for $g_3$. These values are then expanded utilizing a Gaussian probabilistic distribution $D_k$:
	\begin{equation}\label{MBTReq}
		D_k(x,g_k)=\frac{1}{\sigma_k\sqrt{2\pi}}e^{-\frac{(x-g_k)^2}{2\sigma_k^2}}
	\end{equation}
	where $\sigma_k$ is the standard devation of the gaussian kernel. The range, $x$, must contain the values of $g_k$. If the distance between two atoms is large, their contribution to the description of the structure must be reduced. Exponential weighting functions perform these conditions:
	\begin{equation}
		f_1^{Z_1}(x)=\sum_{l}^{|Z_1|}w_1^lD_1^l(x)
	\end{equation}
	\begin{equation}
		f_2^{Z_1,Z_2}(x)=\sum_{l}^{|Z_1|}\sum_{m}^{|Z_2|}w_2^{l,m}D_2^{l,m}(x)
	\end{equation}
	\begin{equation}
		f_3^{Z_1,Z_2,Z_3}(x)=\sum_{l}^{|Z_1|}\sum_{m}^{|Z_2|}\sum_{n}^{|Z_3|}w_3^{l,m,n}D_3^{l,m,n}(x)
	\end{equation}
	There is no need to use special weighting functions for $k=1$. But for $k=2$ and $k=3$ the weight functions are as follows 
	\begin{equation}
		w_2^{l,m}=e^{-s_k|R_l-R_m|}
	\end{equation}
	\begin{equation}
		w_3^{l,m,n}=e^{-s_k|R_l-R_m|+|R_m-R_n|+|R_l-R_n|}
	\end{equation}
	The $s_k$ parameter is used to set the cutoff distance. The values of $\sigma_k$ for $k=1,2,3$, as well as the range selected for $x$, to produce the MBTR from the structures with respect to Eq.\ref{MBTReq} are shown in table \ref{tabMBTRH}. 
	These values again are investigated through the grid search process.
	In the implementation of the Dscribe package, an additional parameter $w_k^{min}$ is also introduced to remove elements less than $w_k^{min}$, table \ref{tabMBTRH} shows the values selected for these parameters. The weight function $w_1=1$ is considered here for one-body functions, so the $s_k$ and $w_k^{min}$ values are not defined for one-body functions. Here, due to the use of the cosine function for 3-body functions, the values for these functions are always between 1 and -1, so the range of x is considered for 3-body functions between 1 and -1.
	\begin{table}[th!]
		\centering
		\caption{\small Hyperparameter values used to generate MBTR descriptors. $\sigma_k$ is the standard deviation of the Gaussian function, and $s_k$ control the exponent of the exponential weights. $w_k^{min}$ is the minimum threshold for MBTR elements. These parameters are obtained through gridsearch process that for Co$_2$Ni$_2$Nb$_2$O$_9$(FM), Co$_2$Ni$_2$Nb$_2$O$_9$(AF), MnFe$_2$O$_4$, and Ca$_{8.63}$Sb$_{10}$Sr$_{2.37}$ trained on 100 structures, and for Cr$_{0.2}$Mn$_{0.4}$Ni$_{1.2}$Zr trained on 90, and for NaCaNi$_2$F$_7$ trained on 40 structures.}
		\resizebox{\columnwidth}{!}{%
		\begin{tabular}{c||ccccccc} \hline\hline
			compound&$k$&$n_k$&$\sigma_k$&$s_{k}$&$w_k^{min}$&$\lambda$&$\sigma$\\ \hline
			\multirow{3}{*}{NaCaNi$_2$F$_7$}&1&88&0.01&-&-&\multirow{3}{*}{$3.16\times10^{-15}$}&\multirow{3}{*}{-1196.63369}\\
			&2&88&0.01&0.5&$10^{-3}$\\
			&3&88&0.01&0.5&$10^{-4}$\\\hline
			\multirow{3}{*}{Co$_2$Ni$_2$Nb$_2$O$_9$(AF)}&1&24&0.15&-&-&\multirow{3}{*}{$5.75\times10^{-14}$}&\multirow{3}{*}{-1809.03065}\\
			&2&60&0.15&0.51&0.0071\\
			&3&54&0.03&0.57&$10^{-4}$\\\hline
			\multirow{3}{*}{Co$_2$Ni$_2$Nb$_2$O$_9$(FM)}&1&64&0.16&-&-&\multirow{3}{*}{$5.01\times10^{-15}$}&\multirow{3}{*}{-1809.03218}\\
			&2&82&0.32&0.99&0.0021\\
			&3&82&0.001&0.5&0.0011\\\hline
			\multirow{3}{*}{Ca$_{8.63}$Sb$_{10}$Sr$_{2.37}$}&1&84&0.01&-&-&\multirow{3}{*}{$2.51\times10^{-15}$}&\multirow{3}{*}{-2658.87585}\\
			&2&84&0.01&0.5&$10^{-3}$\\
			&3&84&0.01&0.5&$10^{-4}$\\\hline
			\multirow{3}{*}{MnFe$_2$O$_4$}&1&64&0.23&-&-&\multirow{3}{*}{$3.7152\times10^{-14}$}&\multirow{3}{*}{-840.49802}\\
			&2&83&0.39&0.57&0.0001\\
			&3&53&0.16&0.78&0.0001\\\hline
			\multirow{3}{*}{Cr$_{0.2}$Mn$_{0.6}$Ni$_{1.2}$Zr}&1&76&0.31&-&-&\multirow{3}{*}{$1.30\times10^{-11}$}&\multirow{3}{*}{-681.90051}\\
			&2&64&0.43&0.62&0.0061\\
			&3&67&0.0001&0.5&0.0021\\\hline
		\end{tabular}}
		\label{tabMBTRH}
	\end{table}
	\subsection{Atom-Centered Symmetry Function (ACSF)}
	In ACSF, the atomic structures are first divided into atomic environments, where the atomic environments are spheres with radius $R_{cut}$ and the center of the atom $i$ in the system. Each atomic environment is then encoded by symmetric functions. For the radial degree of freedom the 2-body symmetric functions $G_i^1$,$G_i^2$,$G_i^3$ are used and for the angular degree of freedom the 3-body symmetric functions $G_i^4$,$G_i^5$ are used. A set of symmetric functions introduced in~\cite{behler2015constructing} as follows:
	\begin{equation}
		G_i^1=\sum_{j}f_c(R_{ij})
	\end{equation}
	\begin{equation}
		G_i^2=\sum_{j}e^{-\eta(R_{ij}-R_s)^2}f_c(R_{ij})
	\end{equation}
	\begin{equation}
		G_i^3=\sum_{j}\cos(\kappa R_{ij})f_c(R_{ij})
	\end{equation}
\begin{multline*}
			G_i^4=2^{1-\zeta}\sum_{j,k\neq i}^{all}(1+\lambda \cos\theta_{ijk})^\zeta \\
			e^{-\eta (R_{ij}^2+R_{ik}^2+R_{jk}^2)} f_c(R_{ij})f_c(R_{ik})f_c(R_{jk})
\end{multline*}
\begin{multline*}
			G_i^5=2^{1-\zeta}\sum_{j,k\neq i}^{all}(1+\lambda \cos\theta_{ijk})^\zeta \\
			e^{-\eta (R_{ij}^2+R_{ik}^2)} f_c(R_{ij})f_c(R_{ik})
\end{multline*}

$G_1$ is the sum of the cutoff functions with respect to the atoms around the central atom $i$. $G_2$ calculates the atomic density around the atom $i$, which must be multiplied by the cutoff function to be smooth and go to zero in the cutoff radius. Parameter $\eta$ determines the Gaussian width, And the Gaussian center can be shifted with parameter $R_s$. $G^3$ provides a cosine description of the radial degree of freedom of the atomic environment. To achieve a suitable radial description, radial functions with different parameters must be used, for example, $G^2$ with a set of cutoff radius or a set of $\eta$ for each atomic environment. Appropriate cutoff functions have been introduced so far, however, each must satisfy that the values of these functions and their derivatives must be smoothly zeroed at distances close to the cutoff radius $ R_c $. Two types of cutoff functions re introduce in ~\cite{behler2015constructing} as:
	\begin{small}
		\begin{equation}
			f_c^1(R_{ij})=
			\begin{cases}
				0.5.\left[\cos \left(\frac{\pi R_{ij}}{R_c}\right)+1\right]& \text{for} \hskip 0.5cm R_{ij}\leq R_c\\
				0.0&  \text{for} \hskip 0.5cm R_{ij}\geq R_c
			\end{cases}
		\end{equation}
		\begin{equation}
			f_c^2(R_{ij})=
			\begin{cases}
				\tanh^3 \left[1-\frac{R_{ij}}{R_c}\right]& \text{for} \hskip 0.5cm R_{ij}\leq R_c\\
				0.0&  \text{for} \hskip 0.5cm R_{ij}\geq R_c
			\end{cases}
		\end{equation}
	\end{small}
	As the distance between the neighboring atom and the central atom increases, the interaction between the two atoms decreases, as clearly seen in both of the introduced cutoff functions. The value of the cutoff function in the cutoff radius $R_c$ is smoothly reduced to zero. 
	The $G^4$ and $G^5$ functions are introduced to describe the angular degree of freedom of the system, also make sure that the neighboring atoms of the central atom that are outside the cutoff radius are not considered and as the distance between two atoms in a triple increase, their share decreases\cite{behler2011atom}.
	
	\begin{table}[th!]
		\centering
		\caption{\small Hyperparameter values used to generate ACSF descriptors. $\sigma_k$ is the standard deviation of the Gaussian function, and $\lambda$ is the regularization parameter. $r_{cut}$ is the cutoff radius. These parameters are obtained through grid search process that for Co$_2$Ni$_2$Nb$_2$O$_9$(FM), Co$_2$Ni$_2$Nb$_2$O$_9$(AF), MnFe$_2$O$_4$, and Ca$_{8.63}$Sb$_{10}$Sr$_{2.37}$ trained on 100 structures, and for Cr$_{0.2}$Mn$_{0.4}$Ni$_{1.2}$Zr trained on 90, and for NaCaNi$_2$F$_7$ trained on 40 structures.}
		\resizebox{\columnwidth}{!}{%
		\begin{tabular}{c||cccc} \hline\hline
			compound&$\eta$&$r_{cut}$&$\lambda$&$\sigma$\\ \hline
			\multirow{6}{*}{NaCaNi$_2$F$_7$}&0.001571&\multirow{6}{*}{16.39}&\multirow{6}{*}{$3.16\times10^{-15}$}&\multirow{6}{*}{-1195.6526}\\
			&0.004575\\
			&0.010293\\
			&0.014767\\
			&0.027074\\
			&0.036220\\\hline
			\multirow{6}{*}{Co$_2$Ni$_2$Nb$_2$O$_9$(AF)}&$6.72\times10^{-5}$&\multirow{6}{*}{4.708}&\multirow{6}{*}{$1.905\times10^{-14}$}&\multirow{6}{*}{-1808.9842}\\
			&0.002836\\
			&0.003328\\
			&0.006864\\
			&0.021921\\
			&0.328821\\\hline
			\multirow{6}{*}{Co$_2$Ni$_2$Nb$_2$O$_9$(FM)}&0.000311&\multirow{6}{*}{6}&\multirow{6}{*}{$6.02\times10^{-14}$}&\multirow{6}{*}{-1809.0154}\\
			&0.000620\\
			&0.000850\\
			&0.001200\\
			&0.004370\\\hline
			\multirow{6}{*}{Ca$_{8.63}$Sb$_{10}$Sr$_{2.37}$}&0.079876&\multirow{6}{*}{14.75}&\multirow{6}{*}{$2.45\times10^{-15}$}&\multirow{6}{*}{-2643.4584}\\
			&0.219659\\
			&0.339786\\
			&0.487519\\
			&0.540577\\
			&0.661157\\\hline
			\multirow{6}{*}{MnFe$_2$O$_4$}&0.009603&\multirow{6}{*}{15.1}&\multirow{6}{*}{$9.77\times10^{-14}$}&\multirow{6}{*}{-843.4154}\\
			&0.011727\\
			&0.014159\\
			&0.033614\\
			&0.033662\\
			&0.256198\\\hline
			\multirow{6}{*}{Cr$_{0.2}$Mn$_{0.6}$Ni$_{1.2}$Zr}&0.025142&\multirow{6}{*}{15.16}&\multirow{6}{*}{$9.12\times10^{-10}$}&\multirow{6}{*}{-683.8965}\\
			&0.040146\\
			&0.128268\\
			&0.171522\\
			&0.253443\\
			&0.766330\\\hline
		\end{tabular}}
		\label{tabACSF}
	\end{table}
	
	\subsection{Crystal Graph Convolutional Neural Network (CGCNN)}
	Table \ref{cgcnn} shows suitable values for maximum distance between atoms (radius), the number of hidden layers of fully-connected neural network (n-h), the maximum number of convolutions (n-convs), the maximum number of neighbors around each central atom (max-num-Nbr), and step parameter that control dimensions of the feature vectors related to edges.
	\begin{table}
		\centering
		\caption{{\small Hyperparameter values used CGCNN. Radius is the maximum distance between atoms, n-h is the number of hidden layers of the fully-connected neural network. n-convs is the maximum number of convolutions, max-num-nbr is the maximum number of atoms that surrounded central atoms, and step is a parameter to control the edge feature vector dimension. These parameters are obtained through the grid search process that is trained on 25\% of the data.}}
		\resizebox{\columnwidth}{!}{%
		\begin{tabular}{|c|c|c|c|c|c|}
			\hline compound&radius(\AA)&n-h&n-convs&max-nub-nbr&step\\\hline
			{NaCaNi$_2$F$_7$}&10&1&20&20&0.4\\
			{Co$_2$Ni$_2$Nb$_2$O$_9$(AF)}&16&3&19&15&0.1\\
			{Co$_2$Ni$_2$Nb$_2$O$_9$(FM)}&12&2&15&15&0.1\\
			{Ca$_{8.63}$Sb$_{10}$Sr$_{2.37}$}&9&2&18&18&0.5\\
			{MnFe$_2$O$_4$}&10&2&21&17&0.3\\
			{Cr$_{0.2}$Mn$_{0.6}$Ni$_{1.2}$Zr}&18&2&8&19&0.3\\\hline
		\end{tabular}\label{cgcnn}}
	\end{table}
	
	\newpage
	
	\bibliographystyle{apsrev4-1}
	\bibliography{refrence.bib} 
\end{document}